\documentclass[aps,pre,preprint,groupedaddress]{revtex4}
\usepackage{graphicx}
\usepackage[fleqn]{amsmath}
\usepackage{bm}
\usepackage{epsf,epsfig,graphics}
\usepackage{float}
\usepackage{makeidx}
\usepackage{latexsym}
\usepackage{amssymb}
\usepackage{mathrsfs}
\usepackage{amsbsy}
\usepackage{subfigure}

\newcommand{\figws}{8cm}

\newcommand{\figw}{10cm}

\usepackage{color}
\usepackage{morefloats}
\DeclareMathAlphabet\mathbfcal{OMS}{cmsy}{b}{n}

\begin{document}

\title{Phase Transitions in 
 Ehrenfest Urns Model with Interactions: Coexistence of uniform and non-uniform states}
\author{Chi-Ho Cheng\footnote{Email: phcch@cc.ncue.edu.tw}$^1$, Beverly Gemao$^{2,3}$ and Pik-Yin Lai\footnote{Email: pylai@phy.ncu.edu.tw}$^2$} 
\affiliation{$^1$Department of Physics, National Changhua University of Education, Changhua 500, Taiwan, R.O.C.}
\affiliation{$^2$Dept. of Physics and Center for Complex Systems, National Central University, Chung-Li District, Taoyuan City 320, Taiwan, R.O.C.}
\affiliation{$^3$Physics Department, MSU-Iligan Institute of Technology, 9200 Iligan City ,  Philippines}
\date{\today}

\begin{abstract}
A model based on the classic non-interacting Ehrenfest urn model with two-urns is  generalized to $M$ urns  with the introduction of interactions for particles within the same urn.  As the inter-particle interaction strength is varied,  phases of different levels of non-uniformity emerge and their stabilities are calculated analytically. In particular, coexistence of locally stable uniform and non-uniform phases  connected by first-order transition occurs. The phase transition threshold and energy barrier can be derived exactly together with the phase diagram  obtained analytically. These analytic results are further confirmed by Monte Carlo simulations.
\end{abstract} 
\maketitle

\section{Introduction}
 In 1872, when Boltzmann formulated the H-theorem\cite{Boltzmann} to explain how a system approaches  equilibrium from non-equilibrium and the irreversibility associated with the second-law of thermodynamics, it also lead to the microscopic time-reversal and the Poincar\'e recurrence paradoxes\cite{huang}, which were not fully understood at that time. 
Decades later, the Ehrenfest two-urn model\cite{Ehrenfest} was
proposed in 1907 to resolve the paradoxes and clarify the relationship between reversible microscopic dynamics and irreversible thermodynamics.  
The classic Ehrenfest model\cite{Ehrenfest} considered a total of $N$ particles distributed in two urns with each particle in an urn to be chosen randomly and put into the other with equal probability. The Ehrenfest urn model is a simple and tractable  model to understand or illustrate the conceptual foundation of statistical mechanics and the relaxation to equilibrium. This model was solved exactly by Kac\cite{kac} and has been often used to demonstrate the second law of thermodynamics and the approach to equilibrium.
  
 Later on, the Ehrenfest model was generalized
to the case of unbalanced  jumping rates between the two urns \cite{siegert,klein}. The two-urn Ehrenfest model was subsequently  extended
 to  multi-urn systems\cite{iglehart,kao2003,kao2004,nagler2005} to investigate the associated non-equilibrium steady-states. Its various generalizations  have been applied to investigate a variety of non-equilibrium phenomena. 
 The continuous time limit of the
evolution of the  population probability state lead to a linear Fokker-
Planck equation\cite{kac,[18]} which was further modified to incorporate the nonlinear contribution 
\cite{[19],[20],[21]}, which is motivated by the processes associated with
anomalous-diffusion phenomena\cite{[22],[23],[24]}. The associated generalized H-theorem for the nonlinear Fokker-Planck equation was also studied\cite{[25],[26],[27],[28],meerson}.
 However, most of such
generalization is non-interacting, or the inclusion of interaction is phenomenological and not explicit.
Until recently, the two-urn Ehrenfest model was extended to included particle interactions inside an urn\cite{chc2017}.  In the two-urn Ehrenfest model with interaction, particles can interact with all other particles inside the same urns, but particles belonging to different urns do not interact. In addition,   
a  jumping rate (asymmetric in general) from one urn to another is introduced, which is independent of the particle interaction. The system can exhibit interesting phase transitions and the Poincar\'e cycle and relaxation times can be calculated\cite{chc2017}.

 In this paper, we extend the interacting Ehrenfest model to $M$ urns ($M>2$). In particular, we focus on the equilibrium case when detailed balance can be achieved.
A possible application for the present  equilibrium  model and its generalization is the optimization  in partitioning problem\cite{Lai87,Lai88}, such as distributing a fixed amount of total  resource to $M$ locations with a certain cost to be minimized. 
  The equilibrium phase behavior of the model is rather rich and can be investigated in detail. Analytic and exact results are derived for the conditions of the emergence of coexistence of uniform or non-uniform phases and the associated first-order phase transition and energy barrier. Monte Carlo simulations are also performed to verified our theoretical findings.

\section{The M-urns  model with Interactions}
The two-urn interacting model in \cite{chc2017} is extended to the case of $M$-urns.  Similar to the two-urn case\cite{chc2017}, $N$ particles are distributed into the $M$ urns ($M\geqslant 3$ is considered in this paper). Pairwise all-to-all interaction is  introduced only for particles in the same urn and particles in different urns
do not interact. Besides particle interactions,  direct jumping rates is further
introduced between a pair of urns.  In general these jump rates can be asymmetric (unbalanced) and the system is non-equilibrium with  non-zero net particle fluxes. On the other hand, if the particles in any urn are free to make transitions back and forth with another urn with balanced jump rates such that detailed balance is obeyed, the system can achieve   an equilibrium state. In this paper, we will focus on  such an equilibrium situation and the associated phase transition.

The energy or Hamiltonian of the interacting particles in the urns are given by
\begin{equation}
\beta{\cal H}=\frac{1}{2N}\sum_{i=1}^M g_i n_i(n_i-1),\label{Ham}
\end{equation}
where $\beta\equiv 1/(k_B T)$ is the inverse temperature and $g_i$ is the pair-wise interaction (in unit of $k_B T$) of the particles inside the $i^{th}$ urn.
The urns can be thought of as arranged in some periodic lattice, such as a one-dimensional ring,  a completely connected network, or in any undirected network such that the jump rates between neighboring urns are balanced. Under such conditions, with suitable choice of transition dynamics, such as the Metropolis rule, detailed balance is obeyed and the system can achieve thermal equilibrium  with the equilibrium population distribution in the urns being Boltzmann, given by
\begin{equation}
\rho_{eq}({\vec n})\propto \frac{N!}{\prod_{i=1}^M n_i !} e^{-\beta {\cal H}}\propto \frac{N!}{\prod_{i=1}^M n_i !} e^{-\frac{1}{2N}\sum_{i=1}^M g_i n_i(n_i-1)},
\end{equation}
where ${\vec n}\equiv (n_1,\cdots, n_M)^\intercal$.
The fraction of particles in the $i^{th}$ urn is denoted by $x_i$, with the constraint $\sum_{i=1}^Mx_i=1$.
In the large $N\to \infty$ limit, using Stirling approximation and with the fraction $x_i\equiv \frac{n_i}{N}$,  ${\vec x}\equiv (x_1,\cdots, x_{M-1})^\intercal$, and $x_M=1-x_1-x_2-\cdots - x_{M-1}$, one has
\begin{eqnarray}
\rho_{eq}({\vec x})&=&{\cal N}\frac{e^{Nf({\vec x})}}{\sqrt{\prod_{i=1}^M x_i}}, \quad \text{where}\\
f({\vec x})
&=&
 -\sum_{i=1}^{M-1} \left(x_i\ln x_i+{g_i\over 2}x_i^2\right)
 -\left(1-\sum_{i=1}^{M-1} x_i\right)\ln \left( 1-\sum_{i=1}^{M-1} x_i\right)-{g_M\over 2}\left( 1-\sum_{i=1}^{M-1} x_i\right)^2\label{fx}\\
\hbox{ and }& &\quad  {\cal N}^{-1}\equiv \int_{\sum_{i=1}^{M-1}x_i\leqslant 1} \prod_{i=1}^{M-1}dx_i \frac{e^{Nf({\vec x})}}{\sqrt{\prod_{i=1}^M x_i}}.
\end{eqnarray}

The saddle-point, ${\vec x}^*$, is obtained from $\partial f/\partial x_\alpha\rvert_{{\vec x}^*}=0$, $\alpha=1,2,\cdots, M-1$, which leads to the saddle-point equations
\begin{eqnarray}
x_i^*e^{g_ix_i^*}&=&\text{the same constant}, \qquad i=1,2,\cdots, M\label{saddle0}\\
\sum_{i=1}^M x_i^*&=&1, \qquad 0<x_i^*<1.
\end{eqnarray}
Hereafter, unless otherwise stated, we shall consider the case of identical pairwise interactions for all the urns, i.e. $g_i=g$ for $i=1,2,\cdots, M$.

\subsection{Uniform and Non-uniform Equilibrium states}
Since $g_i=g$ for every urn, 
 the uniform solution of ${\vec x^{(0)}}\equiv ({1\over M},\cdots,{1\over M})^\intercal$ is always a saddle-point solution of (\ref{saddle0}).
 In addition $M$ non-uniform saddle-points (related by symmetry) with different values for $x_i^*$'s  can exist.
 Notice that the saddle-points are also the fixed points in the corresponding dynamical system which describes the general non-equilibrium physics of the system.
Since the function $xe^{gx}$ is monotonic increasing in the domain $0\leqslant x\leqslant 1$ for $g\geqslant -1$,  all $x_i^*$ satisfying (\ref{saddle0}) can take one possible value and hence only the uniform state is possible. On the other hand, the function has one peak in  $0\leqslant x\leqslant 1$ for $g<-1$, thus each $x_i^*$ (satisfying (\ref{saddle0}) with $g_i=g$) can take  one of the  two possible values, t allowing the possibility of non-uniform solution in (\ref{saddle0}).
Therefore, if $n$ urns have the fraction being one of the roots, say $x$, the other $M-n$ urns will take the fraction $(1-nx)/(M-n)$.
 Hence one can  derive an equation for the saddle-point(s)
\begin{equation}
x e^{g x}=\frac{(1-n x)} {M-n}e^{\frac{g (1-n x)}{M-n}}, \qquad n=0,1,\cdots, M-1,\label{FP}
\end{equation}
which can also be written as\begin{equation}
{1\over{x}}=n+(M-n)e^{g\frac{Mx-1}{M-n}}.
\label{saddlept}
\end{equation}
$n=0$ represents  uniform distribution (${\vec x^{(0)}}$) of particles in which all $M$ urns have the same fraction of $1/M$. $n$ corresponds to number of urns having the same fraction (say $x$) and the other $M-n$ urns having the same fraction of a different value ($\frac{1-nx}{M-n}$).  Notice that $x=1/M$ is always a solution in (\ref{FP}). It is also easy to see that if $x$ is root of (\ref{FP}) for $n=k\geqslant 1$, then 
$\frac{1-kx}{M-k}$ is also a root for $n=M-k$. Hence $n$ and $M-n$ have the same saddle-points and it is suffice to consider $k=0,1,\cdots, \lfloor{M\over 2}\rfloor$ different states, where $k=0$ is the uniform state and the others $k=1,\cdots, \lfloor{M\over 2}\rfloor$ are non-uniform states with different level of non-uniformity.

\subsection{Saddle-node Bifurcations for  the non-uniform saddle-points}
Now consider first the  simpler case of $M=3$, take for example $n=2$ in (\ref{FP})  with the saddle-point  $(x_1,x_2)=(x,x)$,  where $x$ is given by the roots of 
\begin{equation}
xe^{gx}=(1-2x)e^{g(1-2x)}.\label{e1}
\end{equation}
The stability of the saddle-point is determined by the $2\times 2$ Hessian matrix of $f$ in (\ref{fx})
\begin{equation}
{\bf f''}= -\begin{pmatrix} 2g+\frac{1}{x}+\frac{1}{1-2x} & g+\frac{1}{1-2x} \\ g+\frac{1}{1-2x} &  2g+\frac{1}{x}+\frac{1}{1-2x} \end{pmatrix}.\label{hessf3}
\end{equation}
The saddle-point is stable if $\text{Tr} {\bf f''}<0$ and $\det {\bf f''}>0$, i.e. the real part of the two eigenvalues of $ {\bf f''}$ are both negative.
Using (\ref{hessf3}), one can show that the uniform $(x_1,x_2)=(1/3,1/3)$ saddle-point  is stable for $g>-3$. 
On the other hand, careful examination reveals that $x={1\over 3}$ is always a root in (\ref{e1}) and two smaller roots emerges in a pair (one stable and one unstable) for some negative values of $g<g_c$, characteristics of a saddle-node bifurcation. At the bifurcation point $g_c$ can be determined by the condition of emergence of the pair of (stable and unstable) fixed point together with the condition
\begin{equation}
(xe^{g x})'=((1-2x)e^{g(1-2x)})'.\label{e2}
\end{equation}
$x$ can be eliminated from (\ref{e1}) and (\ref{e2}), then $g_c$ is simply given by the root of the following transcendental equation:
\begin{equation}
1-\sqrt{1+{8\over {3g}}}=2\left(1+\sqrt{1+{8\over {3g}}}\right)\exp[{g\over 4}(1+3\sqrt{1+{8\over {3g}}})],
\end{equation}
which has only a single root of $g_c=-2.74564...$.
In fact, for $g<g_c$ three other stable saddle-points related by symmetry emerge in the $x_1$-$x_2$ phase plane. See Fig. \ref{MCeqm} for the Monte Carlo simulation results displaying $\rho_{eq}(x_1,x_2)$ in the co-existing regime.
 Thus  stable non-uniform equilibrium state exists for $g<g_c$, stable uniform equilibrium state exists for $g>-3$, and bi-stable coexisting equilibrium states of uniform and non-uniform populations occurs for $-3<g<g_c $.

For $M$ urns,  the condition of saddle-node bifurcation is obtained by equating the slopes of lhs and rhs of (\ref{FP}), and using (\ref{FP}) one can derive
\begin{equation}
1+g x+ n x \left(\frac{g}{M-n}+\frac{1}{1-n x}\right)=0.\label{slope}
\end{equation}
(\ref{FP}) and (\ref{slope}) will determine the critical value $g_c(n)$ for the new fixed points  to emerge via saddle-node bifurcations.
For $n=0$, $x=-1/g$ is the solution of (\ref{slope}) and
\begin{equation}
x={1\over {2n}}\left[ 1\pm\sqrt{1+\frac{4n(M-n)}{gM}} \right] \quad \text{ for } n=1,2,\cdots, M-1.\label{xx}
\end{equation}
The threshold values $g_c$ at which new fixed point solutions emerge can be obtained by substituting the solution for $x$ in (\ref{xx}) back to (\ref{FP}) to give $g_c(n=0)= -M$ and for $n>0$, $g_c(n)$ is given by the root of the following transcendental equation
\begin{eqnarray}
1&+&\text{sgn}(\frac{M}{2}-n)\sqrt{1+\frac{4 n (M-n)}{g M}}=
  \frac{n }{M-n}\left(1-\text{sgn}(\frac{M}{2}-n)\right)\sqrt{1+\frac{4 n (M-n)}{g M}}  \nonumber \\
  &\times & \exp \left[\frac{g}{M-n} \left(1-\frac{M}{2
   n}
   \left(1+\text{sgn}(\frac{M}{2}-n)\sqrt{1+\frac{4 n (M-n)}{g M}}
   \right)\right)\right]\\
   & &\text{ where } \text{sgn}(x)\equiv  1 \text{ for } x\geqslant 0 \text{ and } -1 \text{ for } x< 0.
\end{eqnarray}
Notice that $n=k$ and $n=M-k$ ($k \geqslant 1$) have the same $g_c$ and hence the (non-uniform) fixed points emerge together via saddle-node bifurcation.
Furthermore, for $n=M/2$, (i.e. even $M$), the only solution to (\ref{FP}) and (\ref{slope}) is $g_c=-M$ and $x=1/M$, and hence there is no non-uniform fixed point emerge due to saddle-node bifurcation. This special non-uniform fixed point emerges at $g_c=-M$ is via pitchfork bifurcation for even $M$, but careful examination of the  Hessian matrix (\ref{pitchd2f}) reveals that this saddle-point is unstable.
Thus the number of distinct $g_c$'s is $\left\lfloor{{M}\over 2}\right\rfloor$, and  the number of distinct  non-uniform phases (only 1 stable and the rest is unstable as shown below) is $M-1$.

\subsection{Stability for the saddle-points}
The stability condition of the saddle-point ${{\vec x}^*}$is determined by the $(M-1)\times(M-1)$ Hessian matrix $({\bf f''})_{\alpha\beta}\equiv\frac{ \partial ^2f}{\partial x_\alpha\partial x_\beta}\rvert_{{\vec x}^*}$. Direct calculations gives
\begin{equation}
\frac{ \partial ^2f}{\partial x_\alpha\partial x_\beta}\bigg\rvert_{{\vec x}^*}=-(\frac{1}{x_M}+g)-(\frac{1}{x_\alpha}+g)\delta_{\alpha\beta},\qquad x_M\equiv 1-x_1-x_2-\cdots, x_{M-1}.\label{d2f}
\end{equation}
For the uniform saddle-point ${\vec x^{(0)}}\equiv ({1\over M},\cdots,{1\over M})^\intercal$
\begin{equation}
\frac{ \partial ^2f}{\partial x_\alpha\partial x_\beta}\bigg\rvert_{{\vec x^{(0)}}}=-(M+g)(1+\delta_{\alpha\beta})
\end{equation}
whose eigenvalues are $-M(g+M)$ and $-(g+M)$ (with $(M-2)$ degeneracy). Thus the uniform phase  becomes unstable for $g<-M$, i.e. when the inter-particle attraction is strong enough, the uniform phase becomes unstable.

For the first non-uniform saddle-point ${\vec x^{(1)}}\equiv (y,\cdots,y)^\intercal$,      
 $y\neq {1\over M}$ and $y$ is the root of (\ref{FP}) with $n=M-1$ or $n=1$, we have from (\ref{d2f})
\begin{equation}
\frac{ \partial ^2f}{\partial x_\alpha\partial x_\beta}\bigg\rvert_{{\vec x^{(1)}}}=-(\frac{1}{1-(M-1)y}+g)-({1\over y}+g)\delta_{\alpha\beta}
\end{equation}
whose eigenvalues are $-(Mg+\frac{1}{y[1-(M-1)y]})$ and $-(g+{1\over y})$ (with $(M-2)$ degeneracy).

In the case of even $M$,   the non-uniform saddle-point  ${\vec x^{({M\over 2})}}\equiv(y,\cdots,y,\frac{2}{M}-y,\cdots, \frac{2}{M}-y)$ exists, where $y$ is the root in (\ref{FP}) with $n={M\over 2}$, the Hessian matrix is
\begin{equation}
\frac{ \partial ^2f}{\partial x_\alpha\partial x_\beta}\bigg\rvert_{{\vec x^{({M\over 2})}}}=
\begin{cases}
    -(\frac{M}{2-My}+g)-({1\over y}+g)\delta_{\alpha\beta}, & \text{if } \alpha, \beta \leqslant {M\over 2}\\
   -(\frac{M}{2-My}+g)(1+\delta_{\alpha\beta})  , & \text{otherwise}.\label{pitchd2f}
  \end{cases}
\end{equation}
The eigenvalues of (\ref{pitchd2f}) are $-(\frac{M}{2-My}+g)$ (with $({M\over 2}-2)$ degeneracy), $-(g+{1\over y})$ (with $({M\over 2}-1)$ degeneracy), and
$-{1\over 2} [M(\frac{M}{2-My}+g)+ g+{1\over y} \pm \sqrt{M(M-2)(\frac{M}{2-My}+g)^2+(\frac{1}{y}+g)^2}]$. 

\begin{figure}[htbp]
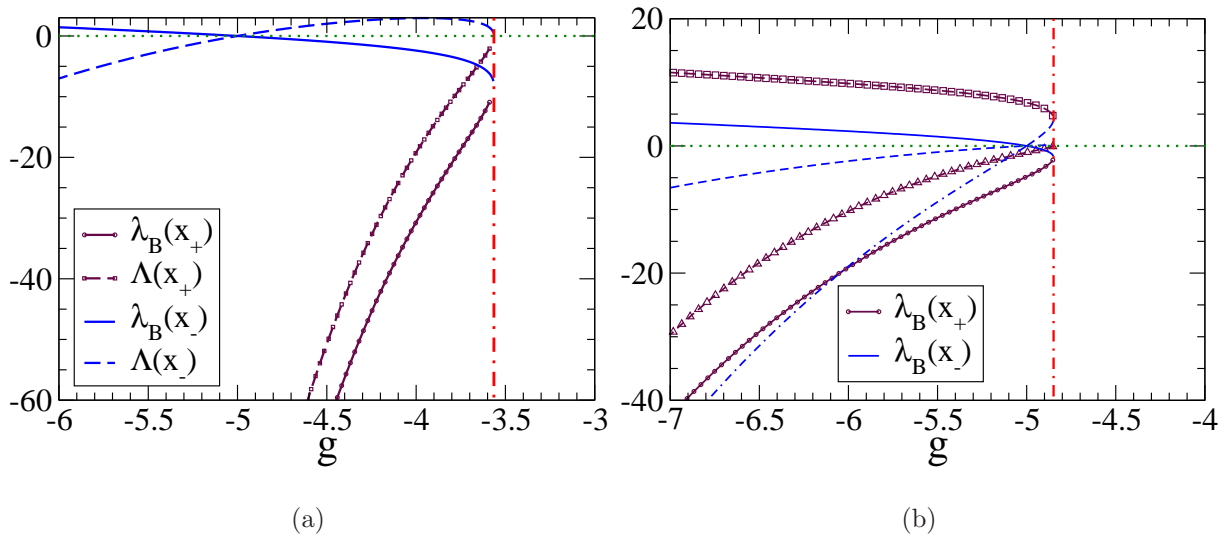

\centering
\subfigure[]{\includegraphics*[width=\figws]{M5k1eigen.eps}}
\subfigure[]{\includegraphics*[width=\figws]{M5k2eigen.eps}}
\caption{ The eigenvalues as a function of $g$ in a $M=5$ system for: (a) the first non-uniform state (solid curve with degeneracy $M-2$ and   a non-degenerate one denoted by the dashed curve).  (b)  The second ($k=2$) non-uniform state (solid curve with degeneracy $M-3$ and   two non-degenerate ones with dashed and dotted curves). The two different colors (brown with symbols and blue without symbol) denote the two non-uniform saddle points $x_+$ and $x_-$ respectively. The value of $g_c$ is marked by a vertical dot-dashed line, and the horizontal dotted line marks the zero value.} \label{eigM5}
\end{figure}

The $k^{th}$ $(k\geqslant 1)$ non-uniform  saddle point can be obtained by putting $n=M-k$ in the saddle-point equation (\ref{saddlept}). Apart from the uniform saddle-point, there are in general two non-uniform root from (\ref{saddlept}), except for $k={M\over 2}$ (even $M$) in which there is only 1 non-uniform root. The eigenvalues of the non-uniform saddle-points can be evaluated as a function of $g$ to reveal the stability of the non-uniform phases (see Appendix for detail calculations). Careful examination of the eigenvalues indicated that only one of the first non-uniform phases is stable and all other non-uniform  ($k>1$) phases always have at least one eigenvalue with a positive real part. Fig. \ref{eigM5} illustrates the results of eigenvalues for the first two non-uniform phases for the case of $M=5$. Only one of the first non-uniform phases has all its eigenvalues negative for all range of $g$, as depicted in Fig. \ref{eigM5}a for the case of $M=5$. For the second non-uniform phase, there is always a positive eigenvalue for both saddle-points in the relevant range of $g$ and hence is an unstable non-uniform phase (see  Fig. \ref{eigM5}b).
It should be noted that one can also employ a dynamical model of the form $\frac{d{\vec x}}{dt}={\vec A}({\vec x})$ whose fixed points are identical with the saddle-point of $f({\vec x})$. And the stability of the fixed points deduced from the Jacobian matrix  $\frac{\partial {\vec A}}{\partial {\vec x}}\rvert_{{\vec x}^*}$ is the same as obtained from the Hessian matrix $\frac{ \partial ^2f}{\partial x_\alpha\partial x_\beta}\rvert_{{\vec x}^*}$.

\section{First-order Phase Transition between uniform and first non-uniform states}
For equilibrium transition  between the coexisting uniform and first non-uniform states as $g$ varies, it is convenient to project onto some line  in the phase space and consider
the projected equilibrium distribution function $\hat{\rho}_{eq}(x)$ parametrized by a single variable $x$. For instance with $M=3$, one can define
\begin{equation}
\hat{\rho}_{eq}(x_1) =\int \rho_{eq}(x_1,x_2)\delta(x_2-x_1)dx_2\propto \frac{e^{Nf( x_1,x_1)}}{\sqrt{x_1(1-2x_1)}}
\end{equation}
which has two maxima at $1/3$ and ${\tilde x}<1/3$. First-order transition occurs at $g=g_t$, which is given by 
\begin{eqnarray}
\frac{\partial}{\partial x}\left( \frac{e^{Nf( x,x)}}{\sqrt{x(1-2x)}} \right)\bigg\rvert_{\tilde x} &=&0\\
\frac{e^{Nf({1\over 3},{1\over 3})}}{{1\over 3}\sqrt{{1\over 3}}} &=& \frac{e^{Nf( {\tilde x},{\tilde x})}}{\sqrt{{\tilde x}(1-2{\tilde x})}}.
\end{eqnarray}
For $N\to \infty$, one can solve to get ${\tilde x}={1\over 6}$ and 
\begin{equation}
g_t=-4\ln 2=-2.77259...
\end{equation}
At $g=g_t$, $\hat{\rho}_{eq}(x)$ has a local minima at $x={1\over 4}$ which in turn gives the energy barrier at the transition, $  \frac{E_b}{N}=\ln 3-\frac{19}{12}\ln 2= 0.00112925...$

\begin{figure}[htbp]
\centering
\subfigure[]{\includegraphics*[width=\figws]{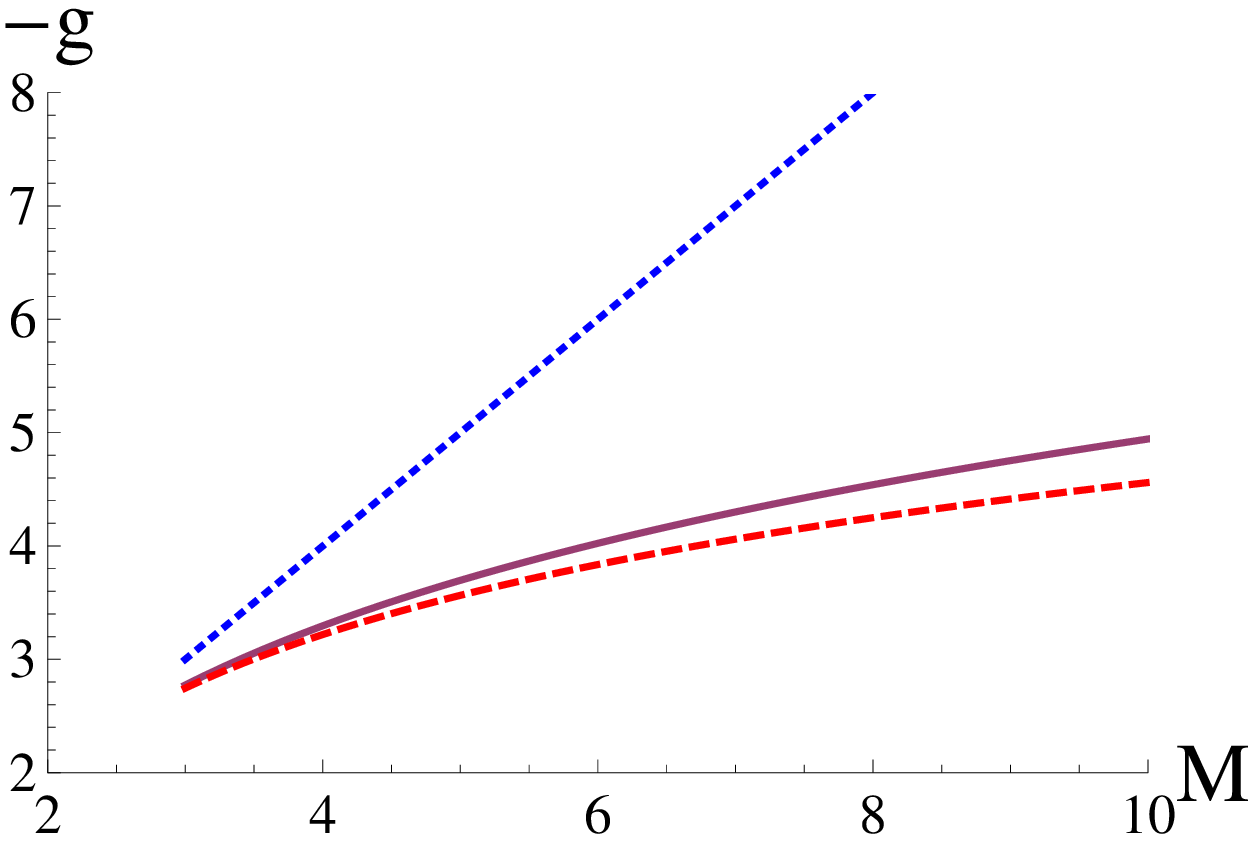}}
\subfigure[]{\includegraphics*[width=\figws]{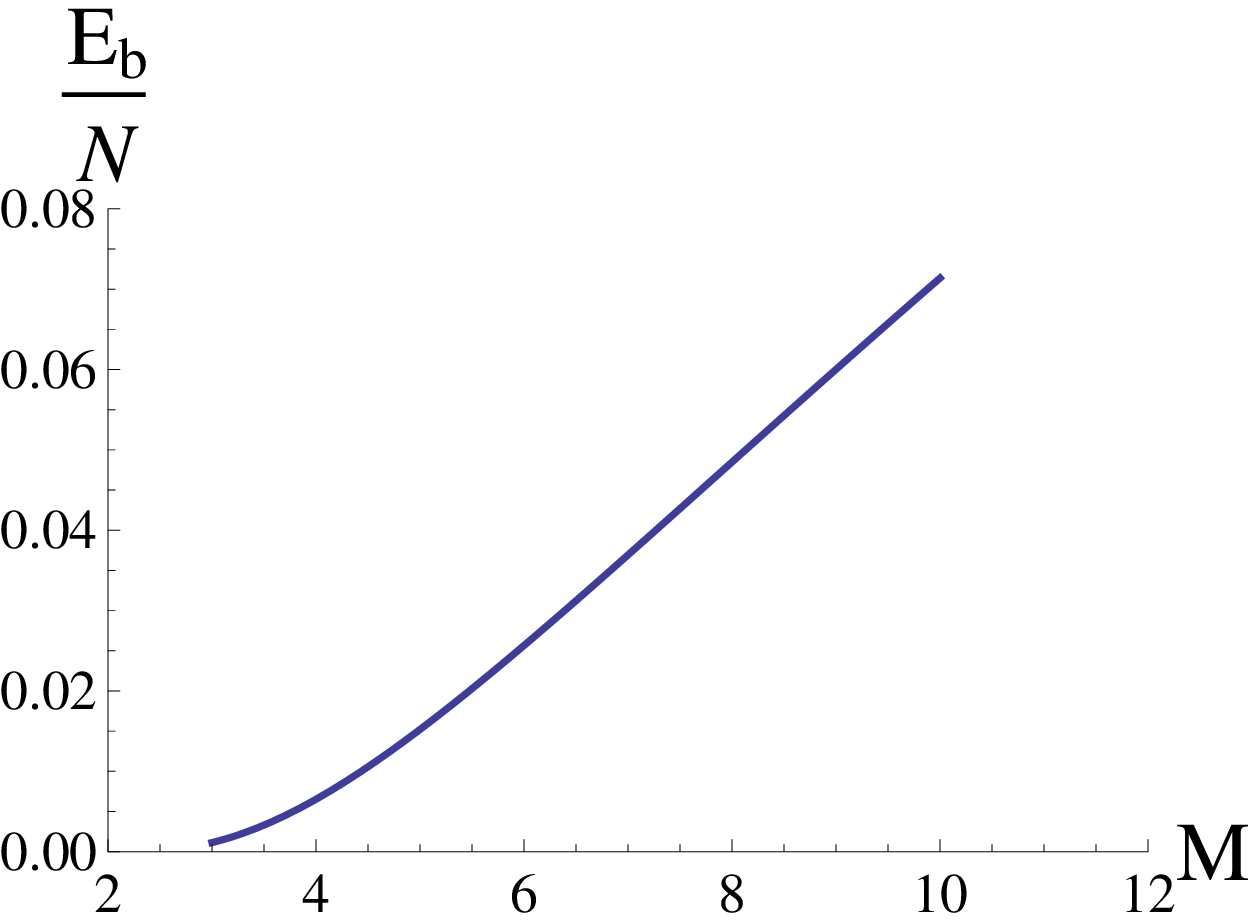}}
\caption{(a) The threshold $g_t$ for first-order transitions between $0\leftrightarrow 1$ (solid curve) and the critical value of $g$ at which the locally stable first non-uniform phase emerges, $g_c(n=1)$ (dashed curve), plotted as a function of $M$. The dotted line denotes $g=-M$ at which the uniform phase becomes unstable. (b) Energy barrier $E_b/N$ vs $M$ for the first-order transitions in (a).} \label{EbvsM}
\end{figure}

In general for $M$ urns, first-order transition occurs at $g=g_t$ which is given by 
\begin{eqnarray}
 \frac{e^{Nf({\vec {\tilde x}})}}{\sqrt{{\tilde x}^{M-1}[1-(M-1){\tilde x}]}}&=&\frac{e^{Nf({\vec x^{(0)}})}}{\sqrt{\frac{1}{M^M}}} \\
\left. \frac{\partial}{\partial x}\left( \frac{e^{Nf({\vec x})}}{\sqrt{x^{M-1}[1-(M-1)x]}} \right)\right|_{\tilde x} &=&0 \qquad {\tilde x}\neq {1\over M}\end{eqnarray}
For $N\to \infty$, one can solve the above equations to get 
\begin{eqnarray}
{\tilde x}&=& \frac{1}{M(M-1)}\\
g_t&=&-\frac{2 (M-1)}{M-2} \ln  (M-1).\label{gt01}
\end{eqnarray}
At $g=g_t$, one can define
$\hat{\rho}_{eq}(x)\equiv \rho_{eq}(x,\cdots,x)$ to characterize the energy barrier.$\hat{\rho}_{eq}(x)$ has a local minima at $x={1\over {2(M-1)}}$ which in turn gives the energy barrier at the transition, 
\begin{equation}
\frac{E^b}{N}=\ln{M\over2} - \frac{3 M - 2}{4 M}\ln(M - 1).\label{EbM}
\end{equation}
Fig. \ref{EbvsM}a plots the first-order transition threshold as a function of $M$, together with $g_c$ at which the first non-uniform phase emerges.  The characteristic energy barrier at the first-order transition as a function of $M$ is shown in Fig. \ref{EbvsM}b.

\section{Equilibrium Phase Diagram}
\begin{figure}[htbp]
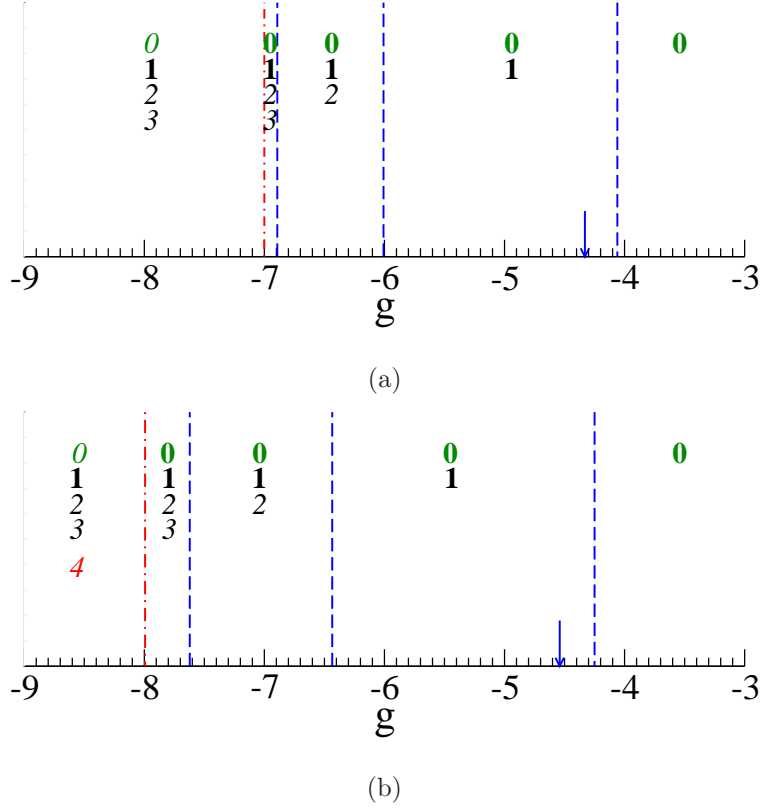

\centering
\subfigure[]{\includegraphics*[width=\figw]{M7phasenew.eps}}
\subfigure[]{\includegraphics*[width=\figw]{M8phasenew.eps}}
\caption{Phase diagrams for (a) $M=7$ and (b) $M=8$ showing  various phases.
Uniform state is denoted by 0 and various non-uniform states of different degree of non-uniformity are denoted by 1, 2, ..., with decreasing non-uniformity. State with a locally stable phase is labeled with a bold font. The most non-uniform ($k=1$) state always has a stable phase. The thermal first-order  phase transition that occurs at $g_t$ is marked by an arrow.} \label{phasediag}
\end{figure}

As the inter-particle attraction becomes stronger ($g$ becomes more  negative),  the system undergoes  a first-order transition from the uniform phase with the emergence of coexisting a locally stable   non-uniform phase at $g=g_c(n=1)$. As $g$ becomes more negative, various other non-uniform phases emerge, albeit not locally stable.
 As $g$ decreases to $g=-M$, the uniform phase becomes unstable and only the stable first non-uniform phase remains. Fig. \ref{phasediag} displays the phase diagrams for odd ($M=7$) and even ($M=8$) values of $M$. The values of $g_c$'s at which various non-uniform phase emerge are calculated analytically. The first-order transition point $g_t$ as given by (\ref{gt01}) is also shown.

\section{Monte Carlo Simulations}
To explicitly verify the theoretical results in previous sections, we carry out Monte Carlo simulations for the $M$ urns system. In the simulation, a total of $N$ ($N$ is an integer multiple of $M$) particles are in the system consisting of $M$ urns and the population of the $i$ urn is denoted by $n_i$. 
The transition probability that  a particle from
the $i$th urn jumps to the $j$th urn is
\begin{equation}
T_{i\to j}=\frac{1}{1+e^{-\frac{g}{N}(n_i-n_j-1)}}.\label{Tij}
\end{equation}
It is easy to see that detailed balance is obeyed with the above transition probability and equilibrium will be achieved after sufficient Monte Carlo steps.

In principle, since we are interested in the equilibrium properties, the urns can  be placed on  any bidirectional network with balanced jump rates between all connected pair of urns and particle transition rules made to satisfy the detailed balance condition such that there is vanishing net particle flux between  every connected pair of urns. 
 
A particle is chosen in random out of all the particles in the $M$ urns (say the $i^{th}$ urn is chosen) and a transition jump is made according to the probability given in (\ref{Tij}). In practice, for the purpose of investigating equilibrium properties, we put the $M$ urns on a one-dimensional ring for simplicity. For urns on a one-dimensional ring, the possible transitions are $j=i\pm 1$ with equal jump rate to the left and right urns.
After some long transient time for equilibration, the populations in each urn or the fraction $x_i(t)$ is recorded for a long sampling time.
Time is in Monte Carlo Steps per particle (MCS/N). One MCS/N means that on average every particle has attempted a jump.

To quantify how non-uniform the state is, we  define
 \begin{equation}
    \psi=\sqrt{\frac{1}{M(M-1)}\sum_{i\neq j}(x_i-x_j)^2}\label{psi}
\end{equation}
as the non-uniformity of the state. $\psi$ can also serve as an order-parameter for the phase transition: $\psi\simeq 0$ for the uniform (disordered) state and $\psi>0$ for the non-uniform (order) state.  $\psi$ can be calculated for states of different degree of non-uniformity (labeled by $k$) as given by (\ref{psik}). One can see that from (\ref{psik}) that $\psi$ decreases monotonically with $k$ and thus $k=1$ is the most non-uniform phase.

\begin{figure}[H]
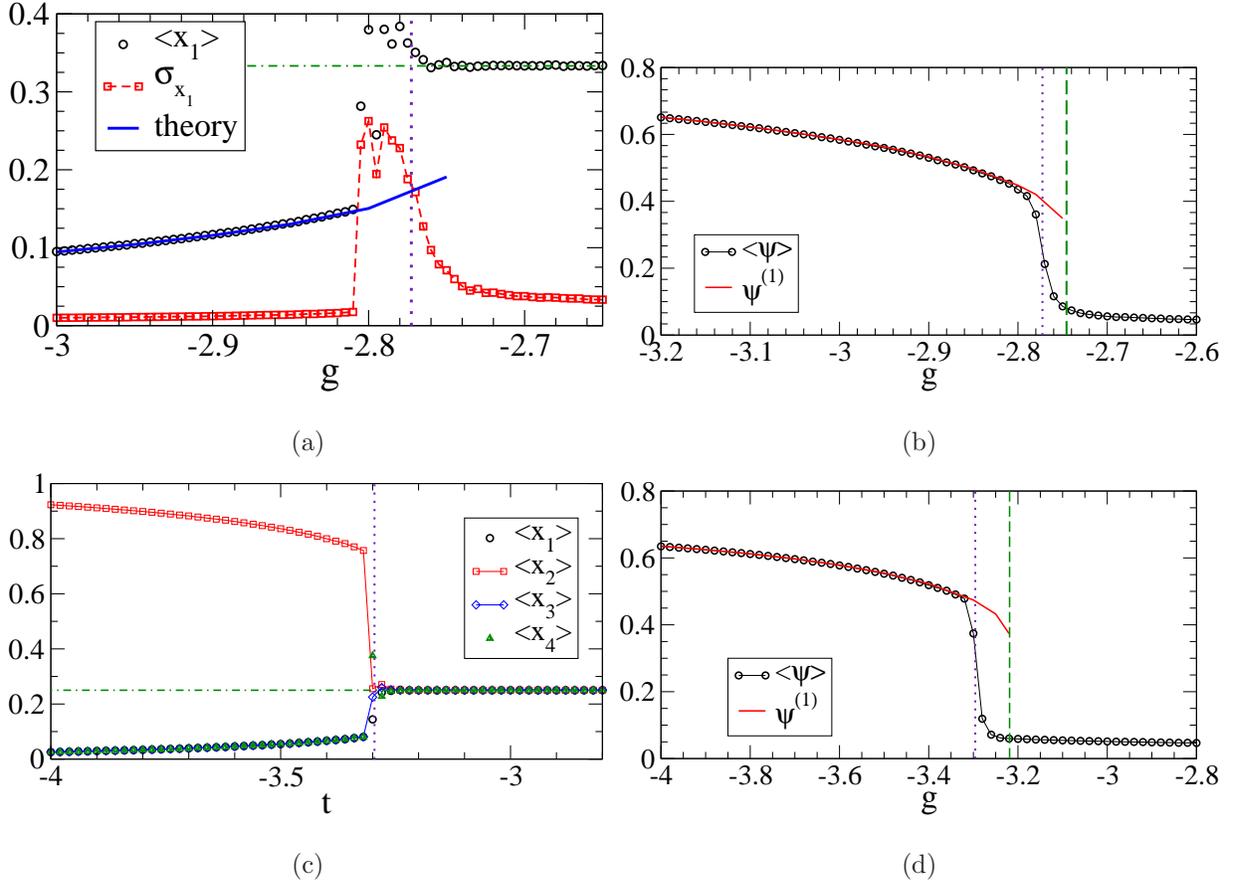

\centering
 \subfigure[]{\includegraphics*[width=\figws]{x1vsgp_5.eps}}
 \subfigure[]{\includegraphics*[width=\figws]{M3nunifvsgp_5.eps}}
\subfigure[]{\includegraphics*[width=\figws]{M4xvsgp_5.eps}}
 \subfigure[]{\includegraphics*[width=\figws]{M4nunifvsgp_5.eps}} 
 \caption{Monte Carlo simulation results  of the urns model for  at equilibrium. 
 (a) 3-urns system with $N=3000$. The mean population and its  fluctuation of one of the urns vs. $g$. The urn with lowest population in the non-uniform state  is chosen. Solid curve is the theoretical value of the mean population which is obtained from the smallest root of the saddle-point equation (\ref{saddlept}). The theoretical first-order transition point is marked by a vertical dotted line. The uniform state with population fraction of ${1\over 3}$ is marked by a horizontal dot-dashed line. $10^5$ MCS/N are used in the sampling. (b) The mean non-uniformity as a function of $g$ in (a). The theoretical  non-uniformity of the first non-uniform state given by (\ref{psik}) is also shown (solid curve). The vertical dashed line marked the theoretical value at which the non-uniform (meta-stable) state emerges. (c)  4-urns system with $N=1000$. The mean populations of the urns vs. $g$.  The theoretical first-order transition point is marked by a vertical dotted line. The uniform state with population fraction of ${1\over 4}$ is marked by a horizontal dot-dashed line. $2\times 10^5$ MCS/N are used in the sampling. (d) The mean non-uniformity as a function of $g$ in (c). The theoretical  non-uniformity of the first non-uniform state given by (\ref{psik}) is also shown (solid curve). The vertical dashed line marked the theoretical value at which the non-uniform (meta-stable) state emerges.}
 \label{MCeqmmeanx}
\end{figure}

 Monte Carlo simulations  for the 3-urn and 4-urn systems as a function of $g$ were carried out results are shown in Fig. \ref{MCeqmmeanx}.  Fig. \ref{MCeqmmeanx}a shows the mean population fraction ($x_1$) of one of the  3 urns  drops from the uniform value of ${1\over 3}$ to a smaller value as the inter-particle attraction increases. The fluctuation of the population fraction, measured by the variance of $x_1$ also shows a peak across the expected first-order transition point. The mean  non-uniformity of the system $\langle \psi\rangle$ also increases as $g$ decreases across the transition. The analytical non-uniformity of the first non-uniform state $ \psi^{(1)}$ is also shown (see Fig. \ref{MCeqmmeanx}b). For inter-particle attraction stronger than $|g_c|$ (marked by vertical dashed line), the first non-uniform phase emerge coexisting with the uniform state.
Fig. \ref{MCeqmmeanx}c shows the mean population fractions of all the urns as a function of $g$ for the 4-urn system at equilibrium. For low attractive strengths, the urns are equally populated with $\langle x_i\rangle\simeq {1\over 4}$. As the inter-particle attraction increases across the predicted first-order transition point ($g_t=-3\ln 3=-3.29584$ from (\ref{gt01}))the populations become inhomogeneous with one urn is more populated and the other three are less but equally populated. The mean  non-uniformity of the system $\langle \psi\rangle$ also shows a sharp rise as shown in Fig. \ref{MCeqmmeanx}d.

\begin{figure}[H]
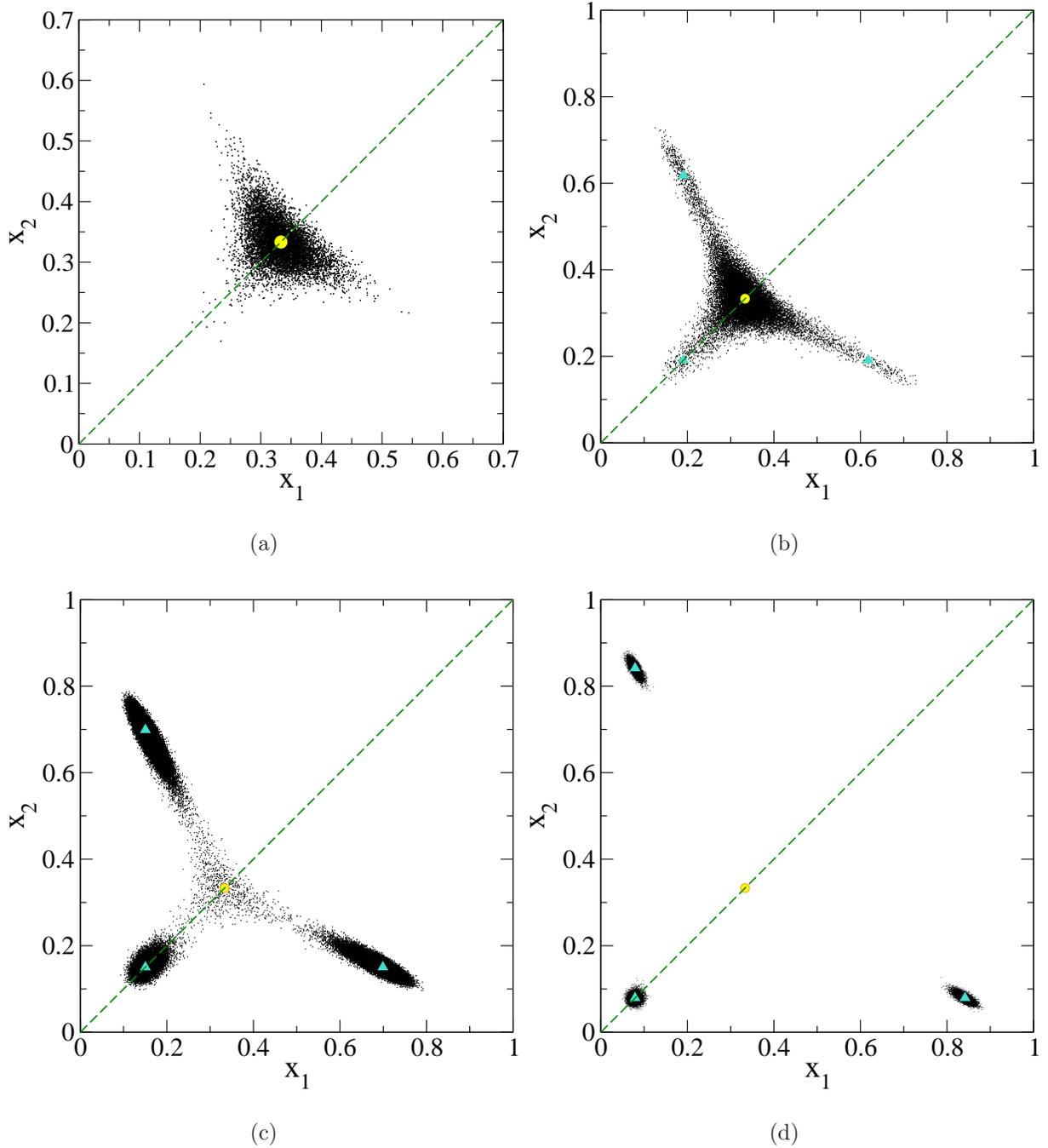

\centering
 \subfigure[]{\includegraphics*[width=\figws]{M3MCx1x2p_5g-2_7b.eps}}
 \subfigure[]{\includegraphics*[width=\figws]{M3MCx1x2p_5g-2_75b.eps}}
 \subfigure[]{\includegraphics*[width=\figws]{M3MCx1x2p_5g-2_8b.eps}}
  \subfigure[]{\includegraphics*[width=\figws]{M3MCx1x2p_5g-3_1.eps}} \caption{Monte Carlo simulation results for the population distribution map of the 3-urns model with $N=3000$  at equilibrium. (a)  $g=-2.7$ in the uniform state. (b) $g=-2.75$  
  and (c) $g=-2.8$ 
    in the co-existing regime. (d)  $g=-3.1$ in the non-uniform state.
     The uniform phase of $x_i={1\over 3}$ is denoted by the yellow filled circle, and the non-uniform phase is denoted by filled triangles.} \label{MCeqm}
\end{figure}

\begin{figure}[H]
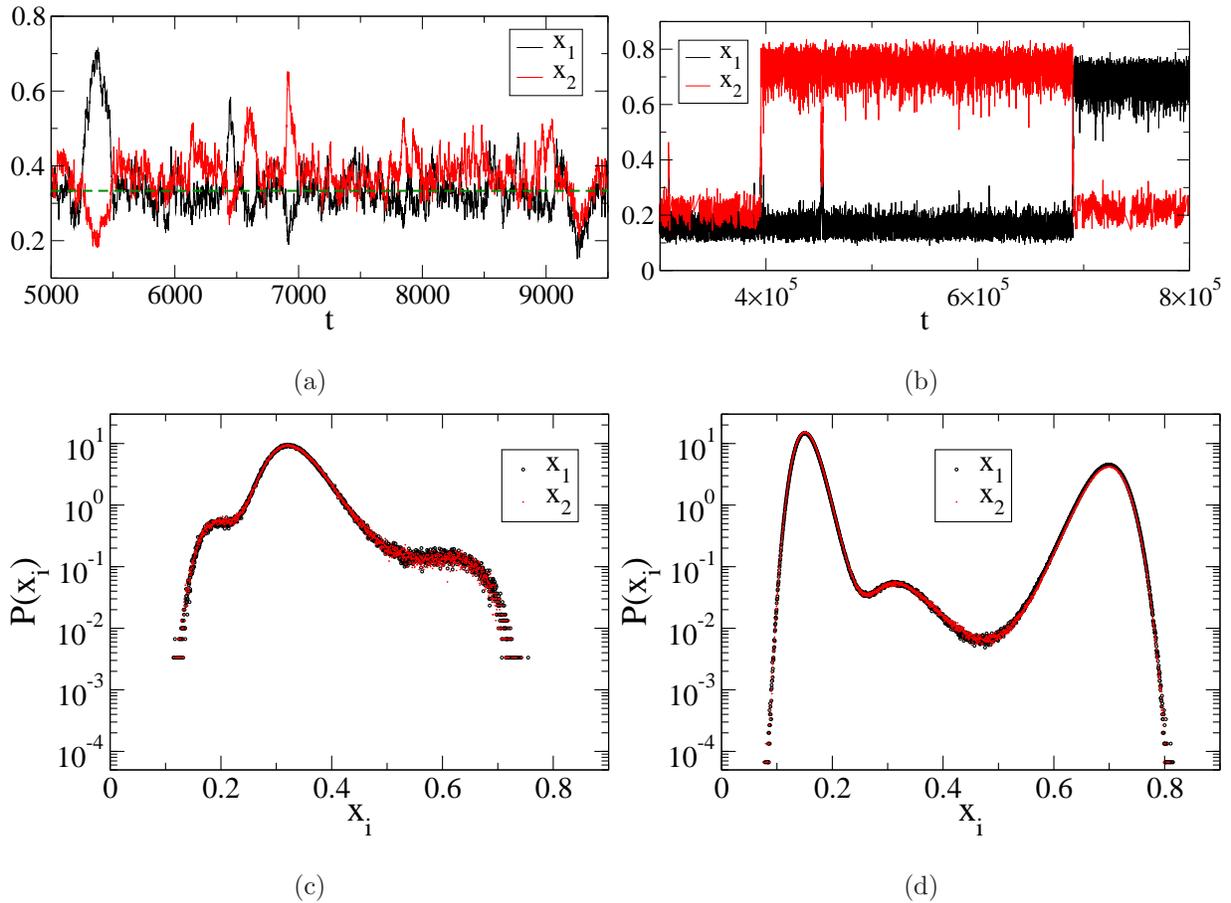

\centering
  \subfigure[]{\includegraphics*[width=\figws]{M3MCxtp_5g-2_75.eps}} 
\subfigure[]{\includegraphics*[width=\figws]{M3MCxtp_5g-2_8.eps}} 
\subfigure[]{\includegraphics*[width=\figws]{M3MCPxp_5g-2_75.eps}} 
\subfigure[]{\includegraphics*[width=\figws]{M3MCPxp_5g-2_8.eps}} 
 \caption{Monte Carlo simulation results for the time course of the populations in the 3-urns model  at equilibrium. $N=3000$. The horizontal dashed line is the uniform state of $x_i={1\over 3}$. Black (darker) curve shows $x_1$ and (grey) curve shows $x_2$. (a)  $g=-2.75$ (b)  $g=-2.8$ in the co-existing regime. Time in Monte Carlo Steps per particle (MCS/N). (c) $P(x_i)$ for the case in (a). $10^6$ MCS/N are used. (d) $P(x_i)$ for the case in (b). $10^8$ MCS/N are used in order to obtain good statistics.} \label{MCeqm2}
\end{figure}

For $M=3$, there are only two independent variables $x_1$ and $x_2$ and the population distribution can be visualized in the two-dimensional density maps shown in Fig. \ref{MCeqm}. For $g>g_c$ the population map has a single peak at the uniform state (Fig. \ref{MCeqm}a), and the non-uniform state emerges and coexist as $g\lesssim g_c$ (Fig. \ref{MCeqm}b). As the inter-particle attraction becomes stronger ($g_t<g<g_c$) the non-uniform population become more significant (Fig. \ref{MCeqm}c). Finally at  $g<g_t$, the uniform state vanishes and only the non-uniform phase remains (Fig. \ref{MCeqm}d).
The time courses of the population fractions of the 3-urn system above and below the first-order transitions are shown in Fig.  \ref{MCeqm2}a and  \ref{MCeqm2}b respectively. For $g\gtrsim g_t$ the system spends most of the time around the uniform state with occasion hopping to the non-uniform meta-stable phases (Fig.  \ref{MCeqm2}a). 
On the other hand for $-3<g<g_t$ , the system is predominantly  in the non-uniform phase but can hop between the degenerate permutation  non-uniform phases in long time scales (Fig.  \ref{MCeqm2}b).  The coexistence of the uniform and non-uniform phases is explicitly spelt out in the distribution functions of ach urns. As shown in Fig.  \ref{MCeqm2}c, the system is dominated by the uniform phase with a prominent peak at $x_i={1\over 3}$, but the two local peaks from the non-uniform phases are clearly seen.  For $g<g_t$, the two peaks of the non-uniform phases grow at the expense of the uniform peak, as shown in Fig.  \ref{MCeqm2}d.

\section{Summary and Outlook}
In this paper, the equilibrium properties of the Ehrenfest $M$-urn model with inter-particle attractions within the same urn is investigated. It is shown that phases of different levels of population non-uniformity can exist, but only the uniform and the most non-uniform phases are local stable. In addition, these two phases can coexist in a range of attraction strengths whose values can be calculated analytically. These two phases are also connected by a first-order transition whose transition interaction strength  (Eq. (\ref{gt01})) and energy barrier (Eq. (\ref{EbM})) can be derived explicitly for arbitrary values of $M$.
For weak $|g|$, the system is in the symmetric (uniform) phase with the same mean population $x_i=1/M$, and for strong $|g|$, the system is the asymmetric phase, and the only stable asymmetric phase is the ($k=1$) most non-uniform state. This first-order phase transition is associated with the breaking of $Z_M$ symmetry as $|g|$ is increased.

The theoretical findings are further verified by Monte Carlo simulations and the agreement is excellent. It is remarkable that as the inter-particle attraction increases, the population changes from the entirely uniform state (in which entropy effects dominates) to the case with the emergence of the locally stable most non-uniform $k=1$ state (in which energy dominates), rather than emerging with a less (or least) non-uniform state. And when the attraction is increased further, less non-uniform states ($k>1$) can emerge, but they are all proved to be unstable. As a result, the most non-uniform state persists and remains stable for $g<g_c(n=1)$ due to the domination of the all-to-all inter-particle attractions within the urn over the entropy effects.
These analytical results and physical picture  can enhance our fundamental understanding of equilibrium phase transitions with multi-phase coexistence. 

 The present model can be extended to  the case  in which the particles can possess internal energy levels. For instance, suppose that the energy spacing of the energy levels at each urn are the same, and the lowest one being zero. Now consider the coupling constant  to be negative so that the particles interaction is attractive. When the temperature is lowered to zero, $g$ approaches to $-\infty$. In this case, inside a urn, the occupation will be dominated by its lowest energy level state. Because of mutual attraction between particles in the same urn, the total number of particles will be located at the lowest energy level of a specific urn. Hence if one generalizes the classical particles to Bosons, also assuming the weak coupling regime and the transition between different urns is classical (no coherence between different urns), then it could possibly lead to Bose condensation in a specific urn.

Here we focused on the equilibrium behavior in which detailed balance is obeyed. But by allowing the jump rates between a pair of urns to be unbalanced, for instance in a one-dimensional ring, the clockwise and anti-clockwise jump rates  are $p$ and  $q$ respectively with $p>q$, then a non-equilibrium state with a net clockwise flux results.  With the particle interaction  explicitly imposed in the model, the interplay of energy and entropy can lead to interesting equilibrium and non-equilibrium phase transitions.
For example, although the  less non-uniform states are found to be unstable,  it  may be  plausible to stabilize them if the inter-urn interactions are introduced in a proper way.
On the other hand, our model can also be extended to other non-equilibrium cases: such as by allowing the particles in the urns be active particles modeled by noise with non-trivial correlations; or the particles are subjected to noises with non-trivial spectrum, then it may lead to additional contributions that could affect the breaking of the ergodicity\cite{Lee,Oliveira} in the broken symmetry non-uniform states. These systems are intrinsically non-equilibrium in nature which is beyond the scope of the present study, but can be investigated in future.

  Finally, we emphasize that the $M$-urn with interaction model  can  serve as a new paradigm model to study various non-trivial equilibrium and non-equilibrium statistical mechanics in a more analytically tractable way, including non-equilibrium steady states or even far from equilibrium situations such as oscillations and even complex spatial-temporal patterns. These are under our current investigations and the results will be presented in  future publications.

\section*{Appendix: Stability calculations for the non-uniform phases}
In this Appendix, we give more details on the definitions of the non-uniform phases and derive their stability conditions. The possible phases are given by the roots of $x$ in  the saddle-point equation (\ref{FP}). As discussed in Sec. II.A, the function $xe^{gx}$ can have at most two distinct values for $0\leqslant x \leqslant 1$, thus at equilibrium the population fractions can only take at most two possible values for a given value of $g$.
Hence we define the $k^{th}$ phase  as the particle distributions such that there are $k$ urns with
the same occupation fraction, say $y$, and the rest ($M-k$) of the urns having the same population fraction, say $x$.
Thus it follows that the  $k^{th}$ and 
 phases are the same and it  suffices 
to consider $k=0,1,\cdots, \lfloor{M\over 2}\rfloor$ possible phases.
 In general $x\neq y$ and $k\neq 0$ for the non-uniform phases, otherwise a uniform phase results.
 It would be more intuitive to rewrite (\ref{FP}) as
\begin{eqnarray}
xe^{gx}=ye^{gy}\label{exey}\\
x=\frac{1-(M-k)y}{k},\label{xy}
\end{eqnarray}
where the relation between $x$ and $y$ in (\ref{xy}) simply follows from the requirement that the sum of all population fractions must be unity.
Since the system possesses permutation symmetry of the $M$ identical urns, one has the freedom to choose the independent coordinates $x_1,x_2,\cdots,x_{M-1}$, i.e. freedom to label the urns using distinct labels.
For actual calculations, we need to choose a convenient labelling. For instance, one can choose the $k^{th}$  phase as given by the $M-1$ component vector
\begin{equation}
{\vec x^{(k)}}= (y,\cdots,y,x \cdots,x)^\intercal \qquad 1\leqslant k \leqslant \left\lfloor{{M}\over 2}\right\rfloor\label{xk}
\end{equation}
whose first $M-k$ components have the same value $y$ (but $y\neq {1\over M}$) and the rest $k-1$ components having the same value of $x=\frac{1-(M-k)y}{k}$.
 The value of $y$ can be solved by substituting (\ref{xy}) into (\ref{exey}) to give 
\begin{equation}
k y=[1-(M-k)y]e^{{g\over k}(1-My)}.\label{kx}
\end{equation}
 The non-uniformity of the $k^{th}$ phase can be computed from (\ref{psi}) to be
\begin{equation}
\psi^{(k)}=\sqrt{\frac{2}{M(M-1)}(\frac{M}{k}-1)} |1-My|.\label{psik}
\end{equation}
In the strong attraction limit, $g\to -\infty$, (\ref{kx}) gives $y\simeq \frac{e^{-|g|/k}}{k}\to 0$ and $ \psi^{(k)}(g\to-\infty)\simeq\sqrt{\frac{2}{M(M-1)}(\frac{M}{k}-1)}\left(1-{M\over k}e^{-|g|/k}\right)\to\sqrt{\frac{2}{M(M-1)}(\frac{M}{k}-1)}$, which is a decreasing function in $k$. Thus the first non-uniform phase ($k=1$) is the most  non-uniform state.

\begin{figure}[H]
\centering
 {\includegraphics*[width=\figws]{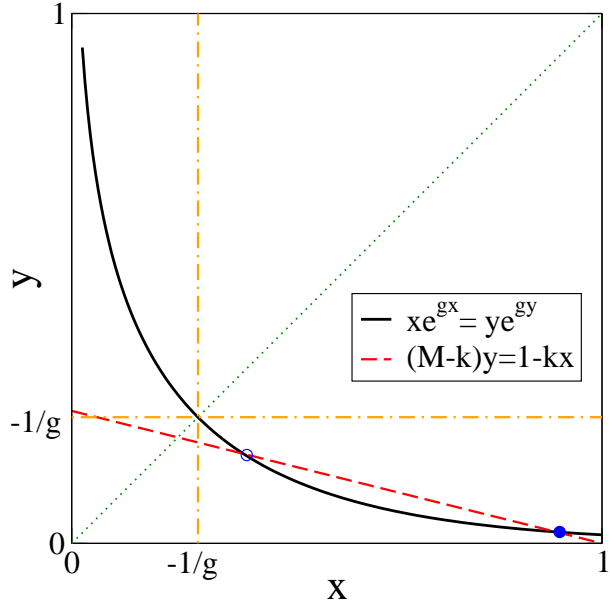}}
\caption{Plots of the curve (\ref{exey}) (for $x\neq y$) and the line (\ref{xy}).
The  two intersections at $x=x_+$ and $x=x_-$ are indicated by filled and open circles respectively. The $x=-{1\over g}$ and $y=-{1\over g}$ are indicated by dot-dashed lines.
The $x=y$ line is indicated by the dotted line.}
\label{exeyfig}
\end{figure}

Apart from the uniform saddle-point ${1\over M}$, there are in general two non-uniform roots of $y$ from (\ref{kx}). More insight can be gained by examining on the $x$-$y$ plane unit square (see Fig. \ref{exeyfig}) in which the intersection of the curve (\ref{exey}) and the line (\ref{xy}) gives the roots for the saddle-points. Consider the case of $x\neq y$ and $k\neq 0$ (non-uniform phases) and $g<-1$, it can be shown\cite{boxes} that the curve (\ref{exey}) always lies outside the square boxes $[0,-{1\over g}] \times [0,-{1\over g}] $ and $[-{1\over g},1] \times [-{1\over g},1] $, and hence the saddle point must satisfy the condition that one of the $x$ or $y$ is $>-{1\over g}$ (but not both), and the other one is $<-{1\over g}$.

\subsection*{Instability for the $k\geqslant 2$ phases}
Here we compute the eigenvalues of Hessian matrix at the $k^{th}$ non-uniform saddle-point which is given by the root of the saddle-point equation (\ref{saddlept}). The stability condition of the $k^{th}$ phase is determined by the $(M-1)\times(M-1)$ Hessian matrix from(\ref{d2f}) and  can be computed by choosing saddle-point ${{\vec x}^{(k)}}$ as in (\ref{xk}) to give
\begin{equation}
\frac{ \partial ^2f}{\partial x_\alpha\partial x_\beta}\bigg\rvert_{{\vec x}^{(k)}}=-\left[\frac{1}{x}+g\right]-\left[\frac{1}{y}+g\right]\delta_{\alpha\beta},\label{d2fk}
\end{equation}
whose eigenvalues can be solved\cite{det} to give (for $2\leqslant k\leqslant \lfloor {M\over 2} \rfloor $)
 \begin{eqnarray}
  \lambda_B&\equiv& -\left({1\over y}+g\right)\qquad  \text{(with $(M-k-1)$ degeneracy)},\label{lamB}\\
  \lambda_A&\equiv& -\left(\frac{1}{x}+g\right) \qquad\text{(with $(k-2)$ degeneracy)}\label{lamA}\\
  & & {1\over 2}\left\lbrace M\lambda_A +\lambda_B \pm \sqrt{( M\lambda_A +\lambda_B)^2-4\lambda_A[(M-k)\lambda_A+k\lambda_B]}\right\rbrace .
 \end{eqnarray}
  These eigenvalues depends on the roots $x$ and $y$ which in turns depend on $g$.
  Now it is easy to see for $k\geqslant 3$, since one of the $x$ or $y$ is $>-{1\over g}$ and hence either $\lambda_A$ or $\lambda_B$ is positive, thus rendering these phases to be always unstable.
  $\lambda_A$ is absent for $k=2$, but we can choose another convenient coordinate such as \begin{equation}
{\vec x^{(k)}}= (x,x,y,\cdots,y)^\intercal\label{xk2}
\end{equation}
and one can compute directly to see that both $\lambda_A$ and $\lambda_B$ are eigenvalues and hence the $k=2$ phases are also unstable.

  \subsection*{Stability and Instability for the $k=1$ phases}
  For $k=1$, it is  convenient to choose the coordinate such that
   \begin{equation}
{\vec x^{(1)}}= (y,\cdots,y)^\intercal\label{xk1}
\end{equation}
and $x=1-(M-1)y$. One can compute directly to find the eigenvalues to be
    $\lambda_B$ (with $(M-2)$ degeneracy) and $\Lambda\equiv (M-1)\lambda_A+\lambda_B=-Mg-\frac{M-1}{x(1-x)}$, where $\lambda_A$ and $\lambda_B$ are given as in (\ref{lamA}) and (\ref{lamB}).
For $g<g_c(n=1)\equiv g_c$ two $k=1$ phases emerges with the corresponding roots $x_+$ and $x_-$ via saddle-node bifurcation, which occurs at $x=x_c$. As $g$ is further decreased, $x_+$ keeps increasing while $x_-$ keeps decreasing. As discussed in previous subsection,  $\lambda_B >0 $ if the root $x < -{1\over g}$  and  $\lambda_B<0$   if $x>-{1\over g}$.
Since the stability also depends on the sign of $\Lambda$, we first find out the conditions that $\Lambda=0$. Vanishing $\Lambda$ occurs for $x$ satisfying $x=\frac{1-x}{M-1}e^{-\frac{1-Mx}{Mx(1-x)}}$. Careful examination of the roots of this equation reveals that there are two roots at $x=x_c$ (the saddle-node bifurcation point at  $g=g_c$) and at
$x=x_-={1\over M}$ (which occurs at $g=-M$).
The eigenvalues $\lambda_B$ and $\Lambda$ evaluated at $x_+$ and $x_-$ determine the stability of these two phases, which are considered for the following two regimes in $g$.

\subsubsection{$-M\leqslant g < g_c$}
We first consider the case of weaker inter-particle attraction $-M\leqslant g < g_c$.
 The condition for saddle-node bifurcation give the relation between $g_c$ and $x_c$ : $M-1=-g_c Mx_c(1-x_c)$, which in turn shows  that the eigenvalue $\Lambda |_{x_c} =0$ at the saddle-node bifurcation point. For $g<g_c$, two roots $x_+ >x_c$ and $x_-<x_c$ emerges, and we will show that the corresponding eigenvalues $\Lambda |_{x_+} <0$ $\Lambda |_{x_-} >0$ in this regime of $g$.
 As $g$ becomes more and more negative, $x_-$ decreases and at $g=-M$,  $x_-={1\over M}$ and the corresponding eigenvalue $\Lambda=0$.  Since $\Lambda|_{x_-}=0$ occurs only at $g=g_c$ and $g=-M$, thus 
$\Lambda|_{x_-}$ does not change sign in the $-M\leqslant g < g_c$ region.  Similarly, $\Lambda |_{x_+} $ will not change sign in the $g<g_c$ region.

We now use perturbation to show that for $g\lesssim g_c$, $\Lambda |_{x_+} <0$ and $\Lambda |_{x_-} >0$. With $g=g_c-\epsilon$ and  writing $x\simeq x_c+\delta$, expanding the saddle-point equation to leading order in $\epsilon$ gives $\delta^2=\frac{2x_c^2(1-x_c)^2(Mx_c-1)}{(M-1)(2x_c-1)}{\epsilon}$. Thus we have
\begin{equation}
\Lambda|_{x_\pm}=\mp \frac{2x_c-1}{x_c^2(1-x_c)^2}(x_\pm-x_c),
\end{equation}
and hence $\Lambda |_{x_+} <0$ and  $\Lambda |_{x_-} >0$ once the saddle-node bifurcation occurs.  Since  $\Lambda|_{x_-}$ does not change sign in the regime of $g$, $x_-$ is unstable. For $x_+$, $\Lambda |_{x_+}$ also does not change sign and remains $<0$,  also the other eigenvalue $\lambda_B<0$ (since $x_+>-{1\over g}$ and  $y_+ <{1\over g}$), thus it is stable.

\subsubsection{$ g <  -M$}
In this case, $x_-<-{1\over g}$ and its eigenvalue $\lambda_B>0$ and this phase is unstable. On the other hand, $x_+$ remains $>-{1\over g}$ and both of its eigenvalues $\lambda_B<0$ and $\Lambda |_{x_+} <0$ ensuring that this is a stable phase.


\begin{acknowledgments}  This work  has been supported by Ministry of Science and Technology of Taiwan under the grant no. 107-2112-M-008-003-MY3, and NCTS of Taiwan.
 \end{acknowledgments}


\begin{thebibliography}{9}
\bibitem{boxes} Without loss of generality, take ${y\over x}<1$, (\ref{exey}) gives $z=e^{gx (1-z)}$ where $z\equiv {y\over x} <1 $. It follows that  $gx <-1$ for the above equation to hold, i.e. $x<-{1\over g}$. Similarly, one can also write  (\ref{exey}) as $z=e^{gy (1-z)}$ where $z\equiv {x\over y} >1 $, now  $gy > -1$ for the above equation to hold, i.e. $y>-{1\over g}$ also. One can repeat the argument by taking ${y\over x}>1$, which leads to the condition of  $x>-{1\over g}$ and  $y<-{1\over g}$. 
Hence the curve (\ref{exey}) for $x\neq y$ lies in region with one of the $x$ or $y$ (exclusively) is $>-{1\over g}$ and the other one $<-{1\over g}$. See Fig. \ref{exeyfig}.
\end{thebibliography}
\end{document}